\documentclass[12pt]{iopart}

\usepackage{graphicx}
\begin{document}
\title[How to modify the van der Waals and Casimir forces]{How to modify
the van der Waals and Casimir forces without change of dielectric permittivity}

\author{G~L~Klimchitskaya${}^1$, U~Mohideen${}^2$ and V~M~Mostepanenko${}^1$
}

\address{${}^1$Central Astronomical Observatory
at Pulkovo of the Russian Academy of Sciences,
St.Petersburg, 196140, Russia}
\address{${}^2$Department of Physics and
Astronomy, University of California, Riverside, California 92521,
USA}

\ead{Umar.Mohideen@ucr.edu}

\begin{abstract}
We propose a new experiment on measuring the Casimir force and its
gradient between an Au-coated sphere and two different plates made
of doped semiconductors. The concentrations of charge carriers in
the plates are chosen slightly below and above the critical
density at which the Mott-Anderson insulator-metal transition
occurs. We calculate changes in the Casimir force and the
Casimir pressure due to the insulator-metal transition using the
standard Lifshitz theory and the phenomenological approach
neglecting the contribution of free charge carriers in
the dielectric permittivity of insulator materials (this approach
was recently supported by the measurement data of several
experiments). It is demonstrated that for the special selection
of semiconductor materials (S- or Se-doped Si, B-doped diamond)
calculation results using both theoretical approaches differ
significantly and the predicted effects are easily detectable
using the existing laboratory setups. In the case that the
prediction of the phenomenological approach is confirmed, this
would open opportunities to modify the van der Waals and
Casimir forces with almost no change of room
temperature dielectric permittivity.
\end{abstract}


\maketitle

\section{Introduction}

The van der Waals and Casimir forces act between closely spaced
material surfaces \cite{1,2}.
These forces are the manifestations of the so-called
{\it dispersion forces} caused by the zero-point and thermal
fluctuations of the electromagnetic field.
The distance range of
 dispersion forces extends from several angstr\"{o}ms to a few
nanometers (the van der Waals regime where the relativistic
retardation is not important) and from a few nanometers to a
few micrometers (the Casimir regime where the retardation effects
contribute more and more as the separation distance increases).
The diverse applications of dispersion forces vary from the physics of
surface and nanostructures \cite{3,4,5,6,7,8,8a,8b} to obtaining
constraints on the predictions of unification theories of
fundamental interactions beyond the Standard Model \cite{9,10,10a}.

The van der Waals and Casimir forces are entirely quantum
phenomena. Theory of these phenomena goes back to the classical
papers by London \cite{11} and Casimir \cite{12}, respectively.
Much work in the theory of dispersion forces was done by
Langreth and his coworkers by developing a density functional
that includes van der Waals interaction \cite{13}.
For plane-parallel layered structures described by the
frequency-dependent dielectric permittivities the exact theory
of the van der Waals and Casimir forces was developed by
Lifshitz \cite{14,15}. In the last few years the main equations
of the Lifshitz theory were generalized for bodies of arbitrary
shape \cite{16,17,18}. This allowed
the development of explicit expressions
for the free energy and force in the experimentally relevant
configurations, such  as a material sphere above a material
plate \cite{19,20}. According to the Lifshitz theory and its
generalizations, in order to modify the van der Waals and Casimir
forces, it is necessary to change the reflection properties of
boundary surfaces over a wide frequency range.

Starting in 1997, a lot of experiments on measuring dispersion
forces between bodies made of different materials have been
performed using modern laboratory techniques
made possible by micro-
and nanotechnology (reviews \cite{21,22,23} contain detailed
description of all experiments with the exception of the most
recent \cite{24,25,26,27,27a}). All these experiments measured
dispersion forces in the retarded regime, i.e., thermal Casimir
forces. Experimental investigation of dispersion forces at
sufficiently high precision leads us to recognize that this
phenomenon is much more complicated than it was
generally believed in the past based on the Lifshitz theory.
The facts suggesting so radical a conclusion are the following.

In a series of dynamic experiments \cite{28,29,30,31} on measuring
the thermal Casimir pressure between two Au plates by means of a
micromachined oscillator it was found that the experimental data
exclude the predictions of the Lifshitz theory
at $T=300\,$K obtained using the
optical data of Au extrapolated to low frequencies by means of
the Drude model (this calculation approach to the thermal
Casimir force was used in 2000 \cite{31a}).
The same data turned out to be consistent with
the predictions of the Lifshitz theory obtained when free charge
carriers are described using the plasma model.
Keeping in mind that the dielectric permittivity of the Drude
model at low frequencies is inversely proportional to the frequency,
as it should be in accordance with the Maxwell equations,
whereas the plasma model neglects relaxation and is approximately
applicable only at sufficiently high frequencies, this result
should be considered as a big surprise. Recently similar
measurements were independently performed by another
technique [i.e., with the help
of an atomic force microscope (AFM) operated in the dynamic
regime \cite{31b,32}] with the same result.
For completeness we note that there are also two experiments
\cite{33,34} on the observation of the Casimir force between
Au-coated spherical lenses of more than 10\,cm radii of
curvature and an Au-coated plate performed by means of a
torsion pendulum. These are not independent measurements of the
Casimir force because they are based on the fit between the
experimental data and theoretical predictions with some fitting
parameters. One of these experiments \cite{33} is in support of
the plasma model and the other one \cite{34} is in favor of
the Drude model (critical analysis of experiments with large
spherical lenses is provided in the literature \cite{35,36,36a}).
Recent experiment using a nanomembrane resonator \cite{36b}
also claims a support of the Drude model, but the comparison
of experiment with theory was shown to be in error \cite{36c}.

Additional evidence comes from measurements \cite{37,38}
of the difference
in the Casimir force between an Au sphere and Si plate
at $T=300\,$K in the
presence and in the absence of a laser pulse on a plate.
The laser pulse led to a 5 orders of magnitude
increase of the free charge carrier
density $n_{\rm fc}$
in the Si sample and respective
change of the dielectric permittivity. The experimental data were
found to be consistent with the Lifshitz theory if the contribution
of free charge carriers to the dielectric permittivity in the absence
of a laser pulse (i.e., in the dark phase when Si is a
dielectric-type semiconductor) is omitted.
In so doing the Casimir force is determined by the contribution of
bound electrons alone.
If the free charge carriers
in the dark phase are included, the theory is found to be inconsistent
with the data. The reason exactly why existing
free charge carriers do
not contribute to the force magnitude remains unclear.
Note that according to the Lifshitz theory bound charge carriers
give a more important contribution to the van der Waals and Casimir
forces at shorter separations, whereas the contribution of free
charge carriers increases with increasing separation.

A similar situation was observed in measurements \cite{39} of the
thermal Casimir-Polder force between ${}^{87}$Rb atoms belonging
to the Bose-Einstein condensate and SiO${}_2$ plate.
The measurement data were found \cite{39} to be consistent with
the
predictions of the Lifshitz theory if dc conductivity of the plate
is omitted. If, however, the dc conductivity of SiO${}_2$ is
included in computations, the obtained theoretical results turn
out to be in disagreement with the data \cite{40}.

Another surprise was from the measurements \cite{26,27,43}
of the thermal Casimir force at $T=300\,$K
between an Au-coated sphere and an indium
tin oxide (ITO) film deposited on a quartz substrate.
At room temperature ITO is a good conductor at low
frequencies, but is transparent to visible and near infrared
light. Based on these properties, ITO was considered \cite{44}
as a very prospective material in experiments on measuring
the Casimir force which require good dc conductivity to correct
for residual electrostatic forces.
Measurements of the gradient of the Casimir
force between an Au sphere and an ITO plate \cite{45,46}
demonstrated that it is 40\%--50\% smaller than between an Au
sphere and an Au plate. It was shown \cite{26,27,43} that
after the UV treatment of an ITO plate the magnitude of the
Casimir force further decreases from 21\% to 35\% depending
on separation. Surprisingly, this decrease is not accompanied
by respective changes in the dielectric permittivity of
ITO which could explain the change in the force based on
the Lifshitz theory. To bring the data into agreement with
theory it was necessary to omit the contribution of free charge
carriers in the UV-treated ITO film. It was hypothesized
\cite{26,27,43} that the UV treatment caused a Mott-Anderson
transition in ITO from a metal to an insulating state without
significant changes in the optical and electrical properties
at room temperature. Then the observed agreement of the data
with the Lifshitz theory when the contribution of free charge
carriers for the UV-treated sample is omitted becomes compartible
with the results of other experiments \cite{37,38,39,40}
discussed above. It should be noted also that the Lifshitz
theory taking into account the static conductivity for dielectric
materials or the relaxation properties of
conduction electrons for metals with perfect
crystal lattices was proved \cite{2,21,44,47} to violate the
third law of thermodynamics (the Nernst heat theorem).
In fact the problems arising in the Lifshitz theory for both
metals and dielectrics have generic roots in the foundations
of quantum statistical physics \cite{45a}.

In this paper we propose a new experiment which could provide direct
validation or disproof of the statement that the transition of
a material from insulating to metal state results in a
significant change of the Casimir force even if the dielectric
permittivity remains nearly unchanged. In the optical
modulation experiment \cite{37,38}, where a Si plate was
radiated with laser pulses, the free charge carrier density
$n_{\rm fc}$ in the
absence of a pulse was much smaller than the critical
concentration at which Si becomes metallic.
As a result, the contributions of free charge carriers in the absence
and in the presence of laser light were quite different.

Here we propose that the Casimir force or its gradient should
be compared between an Au sphere and the two plates
made of doped semiconductor one of which has the doping concentration
slightly below and the other one slightly above the critical
value \cite{47a}. This ensures that the dielectric properties of both plates
at room temperature are almost identical \cite{47a} so that minor differences
between them cannot lead to large change in the magnitude of the
Casimir force between the sphere and each of the plates when this
change is calculated using the standard Lifshitz theory.
As a direct check of the fact that the dielectric permittivities
of both plates are almost identical, their optical properties over
a wide frequency region should be investigated by means of
ellipsometry as was done for ITO samples \cite{26,27,43}.
In order to increase the possible change in the Casimir force
when passing
from the first plate to the second, we choose doped semiconductor
with the critical concentration of charge carriers of order
$10^{20}\,\mbox{cm}^{-3}$ (such as boron-doped diamond or
sulphur-doped Si). As demonstrated below, this would ensure the
increase in the magnitudes of the Casimir force or its gradient
up to 30\% when replacing the insulator-type plate with the
metal-type plate if the effect of Mott-Anderson transition on
dispersion forces does occur, as is anticipated based on
previous experiments \cite{26,27,43}. Such a large relative change
predicted allows reliable measurement of the effect under
discussion with the already built and operated laboratory setups.

The paper is organized as follows. In Sec.~2 we present the main
mathematical expressions for the calculation of dispersion
forces and make the choice of optimal materials. Section~3
is devoted to the calculation of the force and force gradient
for an Au sphere interacting with a plate made of S-doped
silicon when the plate material undergoes a transition from
insulating to metal state. In Sec.~4 the same calculation is
performed for a plate made of B-doped diamond. In Sec.~5
the reader will find our conclusions and discussion.

\section{Expressions for the calculation of dispersion forces
and choice of materials}

We propose to measure in succession the Casimir interaction between
an Au-coated sphere and two plates coated with some doped
semiconductor films whose free charge carrier densities
$n_{\rm fc}$ are slightly
below and slightly above the critical value $n_{\rm fc;cr}$ specific
for this semiconductor. Measurements will be performed by means of
an AFM operated either in the static mode \cite{26,27} or in the
dynamic mode in the frequency shift technique \cite{31b,32}.
In the static regime, the directly measured quantity is the
Casimir force $F$ between the sphere and the plate whereas in the
dynamic regime it is the gradient of the Casimir force
$F^{\prime}=\partial F/\partial a$, where $a$ is the separation
distance between the sphere and the plate. The gradient of the
Casimir force between a sphere and a plate can be simply reformulated
into the Casimir pressure between the two parallel plates (one
made of Au and the other one consists in a semiconductor film
deposited on a substrate) using the proximity force
approximation \cite{2,21}
\begin{equation}
P=-\frac{1}{2\pi R}\,\frac{\partial F}{\partial a}=
-\frac{1}{2\pi R}F^{\prime}.
\label{eq1}
\end{equation}
\noindent
Note that the error from the use of the PFA in the sphere-plate geometry
for a sphere and a plate made of real materials was recently shown
\cite{20} to be less than $a/R$, i.e., about 0.1\% for
the experimental parameters. Because of this, one can safely use
this approximation below in computations based on the Lifshitz
theory.

Using the Lifshitz formula for the free energy in the configuration
of two parallel plates and the PFA, the Casimir force between an
Au sphere of radius $R$ and a semiconductor film deposited on a
substrate at temperature $T$ can be written as
\begin{eqnarray}
&&
F(a,T)=k_BTR\sum_{l=0}^{\infty}{\vphantom{\sum}}^{\prime}
\int_{0}^{\infty}k_{\bot}dk_{\bot}
\nonumber \\
&&~~~~\times
\left\{\ln\left[1-r_{\rm TM}^{(1)}(i\xi_l,k_{\bot})
r_{\rm TM}^{(2)}(i\xi_l,k_{\bot})e^{-2aq_l}\right]\right.
\nonumber \\
&&~~~~~~~
+\left.\ln\left[1-r_{\rm TE}^{(1)}(i\xi_l,k_{\bot})
r_{\rm TE}^{(2)}(i\xi_l,k_{\bot})e^{-2aq_l}\right]\right\}.
\label{eq2}
\end{eqnarray}
\noindent
Here, $k_B$ is the Boltzmann constant, $\xi_l=2\pi k_BTl/\hbar$
with $l=0,\,1,\,2,\,\ldots$ are the Matsubara frequencies, the
prime near the summation sign means that the term with $l=0$ is
divided by 2, $k_{\bot}$ is the projection of the wave vector
onto the plate, and $q_l^2=k_{\bot}^2+\xi_l^2/c^2$.
The reflection coefficients of an Au coating modeled as a
semispace, $r_{\rm TM,TE}^{(1)}$, for the transverse magnetic (TM)
and transverse electric (TE) polarizations of the electromagnetic
field are presented in the form
\begin{eqnarray}
&&
r_{\rm TM}^{(1)}(i\xi_l,k_{\bot})=
\frac{\varepsilon_l^{(1)}q_l-k_l^{(1)}}{\varepsilon_l^{(1)}q_l+
k_l^{(1)}},
\nonumber \\
&&
r_{\rm TE}^{(1)}(i\xi_l,k_{\bot})=
\frac{q_l-k_l^{(1)}}{q_l+k_l^{(1)}},
\label{eq3}
\end{eqnarray}
\noindent
where $\varepsilon_l^{(1)}\equiv\varepsilon^{(1)}(i\xi_l)$ is the
dielectric permittivity of Au along the imaginary frequency axis
and ${k_l^{(1)}}^2=k_{\bot}^2+\varepsilon_l^{(1)}\xi_l^2/c^2$.

The reflection coefficients of a doped semiconductor film of
thickness $d$ deposited on a thick substrate plate modeled as a
semispace are given by \cite{2,48}
\begin{equation}
r_{\rm TM}^{(2)}(i\xi_l,k_{\bot})=
\frac{r_{\rm TM}^{(0,-1)}
+r_{\rm TM}^{(-1,-2)}e^{-2k_l^{(-1)}d}}{1+r_{\rm TM}^{(0,-1)}
r_{\rm TM}^{(-1,-2)}e^{-2k_l^{(-1)}d}}
\label{eq4}
\end{equation}
\noindent
and by a similar expression for $r_{\rm TM}^{(2)}(i\xi_l,k_{\bot})$
with the index TM replaced for TE in (\ref{eq4}).
Here,
$r_{\rm TM,TE}^{(n,n^{\prime})}=r_{\rm TM,TE}^{(n,n^{\prime})}(i\xi_l,k_{\bot})$
are the reflection coefficients on a semiconductor film
($n=0,\,n^{\prime}=-1$) and on a substrate
($n=-1,\,n^{\prime}=-2$)
\begin{eqnarray}
&&
r_{\rm TM}^{(n,n^{\prime})}(i\xi_l,k_{\bot})=
\frac{\varepsilon_l^{(n^{\prime})}k_l^{(n)}-
\varepsilon_l^{(n)}k_l^{(n^{\prime})}}{\varepsilon_l^{(n^{\prime})}k_l^{(n)}
+\varepsilon_l^{(n)}k_l^{(n^{\prime})}},
\nonumber \\
&&
r_{\rm TE}^{(n,n^{\prime})}(i\xi_l,k_{\bot})=
\frac{k_l^{(n)}-k_l^{(n^{\prime})}}{k_l^{(n)}+k_l^{(n^{\prime})}},
\label{eq5}
\end{eqnarray}
\noindent
where $\varepsilon_l^{(-1)}\equiv\varepsilon^{(-1)}(i\xi_l)$ and
 $\varepsilon_l^{(-2)}\equiv\varepsilon^{(-2)}(i\xi_l)$ are the
dielectric permittivities of the semiconductor film and the substrate
material, respectively, and $\varepsilon_l^{(0)}=1$ is the
permittivity of the vacuum gap. We also use the notation
 ${k_l^{(n)}}^2=k_{\bot}^2+\varepsilon_l^{(n)}\xi_l^2/c^2$
 with $n=0,\,-1,\,-2$ [note that $k_l^{(0)}=q_l$].

The Casimir pressure at a temperature $T$, which is calculated using
(\ref{eq1}) from the gradient of the Casimir force measured in
the dynamic regime, is given by
\begin{eqnarray}
&&
P(a,T)=-\frac{k_BT}{\pi}\sum_{l=0}^{\infty}{\vphantom{\sum}}^{\prime}
\int_{0}^{\infty}q_lk_{\bot}dk_{\bot}
\nonumber \\
&&~~~~\times
\left\{\left[\frac{e^{2aq_l}}{r_{\rm TM}^{(1)}(i\xi_l,k_{\bot})
r_{\rm TM}^{(2)}(i\xi_l,k_{\bot})}-1\right]^{-1}\right.
\nonumber \\
&&~~~~~~~
+\left.\left[\frac{e^{2aq_l}}{r_{\rm TE}^{(1)}(i\xi_l,k_{\bot})
r_{\rm TE}^{(2)}(i\xi_l,k_{\bot})}-1\right]^{-1}\right\},
\label{eq6}
\end{eqnarray}
\noindent
where the reflection coefficients for our configuration are
already
defined in (\ref{eq3})--(\ref{eq5}).

Equations (\ref{eq2}) and (\ref{eq6}) can be used to calculate the
Casimir force between an Au sphere and a semiconductor film
deposited on a substrate and the Casimir pressure
between an Au plate and
the same film on a substrate. This can be done in two different
ways: by the immediate substitution of the dielectric permittivity
along the imaginary frequency axis, as obtained using the
Kramers-Kronig relation from the measured optical data (this way
is suggested by the standard Lifshitz theory) and by omitting the
contribution of free charge carriers when the semiconductor film
is in an insulating state (this way is suggested by
 the experimental results discussed in Sec.~1).
Below we choose semiconductor materials leading to the largest
difference between the computational results obtained in both
ways.

Note that an experiment similar in spirit to the one proposed here
was performed \cite{22,49} with two P-doped Si plates which
possess, however, radically different densities of free
charge carriers.
For the first plate interacting with an Au sphere the
concentration of free charge carriers was
$n_{\rm fc}^a\approx 1.2\times 10^{16}\,\mbox{cm}^{-3}$, and for the
second $n_{\rm fc}^b\approx 3.2\times 10^{20}\,\mbox{cm}^{-3}$.
Thus, these concentrations were approximately two orders of
magnitude lower, respectively, higher than the critical
concentration for a P-doped Si equal to \cite{50}
$n_{\rm fc;cr}^{\rm Si:P}\approx 3.84\times 10^{18}\,\mbox{cm}^{-3}$.
Keeping in mind the rather high static dielectric permittivity,
$\varepsilon_{{\rm Si},0}\approx 11.7$, the P-doped Si does not
seem a good candidate for the experiment proposed here.
In a similar way, B-doped Si
($n_{\rm fc;cr}^{\rm Si:B}\approx 3.95\times 10^{18}\,\mbox{cm}^{-3}$
\cite{51}) or Sb-doped Ge
($n_{\rm fc;cr}^{\rm Ge:Sb}\approx 1.55\times 10^{17}\,\mbox{cm}^{-3}$
\cite{50}) are not preferable for using in the proposed
experiment.

Another situation holds for a Si doped by S or Se. For a S-doped
Si the critical concentration of free charge carriers falls in
between
$n_{\rm fc}^{\rm Si:S}\approx 1.8\times 10^{20}\,\mbox{cm}^{-3}$,
when this semiconductor is an insulator, and
$n_{\rm fc}^{\rm Si:S}\approx 4.3\times 10^{20}\,\mbox{cm}^{-3}$
when it is of metallic-type \cite{52}.
For a Se-doped Si the critical concentration is confined between
\cite{53}
$n_{\rm fc}^{\rm Si:Se}\approx 1.4\times 10^{20}\,\mbox{cm}^{-3}$
(semiconductor of dielectric-type) and
$n_{\rm fc}^{\rm Si:Se}\approx 4.9\times 10^{20}\,\mbox{cm}^{-3}$
(semiconductor of metallic-type).

One more prospective material for the purposes of the proposed
experiment is boron-doped diamond. For this semiconductor the
dielectric-metal transition occurs at \cite{54}
$n_{\rm fc;cr}^{\rm C:B}=(4.5\pm 0.5)\times 10^{20}\,\mbox{cm}^{-3}$.
Depending on the preparation technology, the static dielectric
permittivity of B-doped C is confined between 5.5 and 10.
This semiconductor, demonstrating many interesting properties
including superconductivity, is a good candidate for use in the
 proposed experiment.

The semiconductor materials discussed above will be used in
computations of Secs.~3 and 4 in order to determine the
feasibility of the experiment under discussion.

\section{Change of dispersion forces due to dielectric to metal
transition in doped silicon}

To perform computations of the Casimir force between a sphere and
a plate and the Casimir pressure between two flat plates we need
the dielectric permittivities of all materials under discussion
along the imaginary frequency axis. They are going to be obtained
from the reflectivity data measured by means of ellipsometry for
the samples used in experiments. For preliminary theoretical
investigation the tabulated optical data for Si can be used
\cite{55Si}. A less than 1\% error in the magnitude of the Casimir
force, as compared with the tables, is given by the following
analytical approximation suggested for high-resistivity Si
\cite{56}:
\begin{equation}
\varepsilon_{\rm Si}(i\xi)=1.035+
\frac{C_{\rm Si}}{1+\frac{\xi^2}{\omega_{\rm Si}^2}},
\label{eq7}
\end{equation}
\noindent
where $\omega_{\rm Si}=6.6\times 10^{15}\,$rad/s and
$C_{\rm Si}=10.73$.
In figure~1 the behavior of $\varepsilon_{\rm Si}$ as a function
of frequency is shown by the grey line.
%%%%%%%%%%%%%%%
\begin{figure*}[h]
\vspace*{-13.cm}
\centerline{\includegraphics{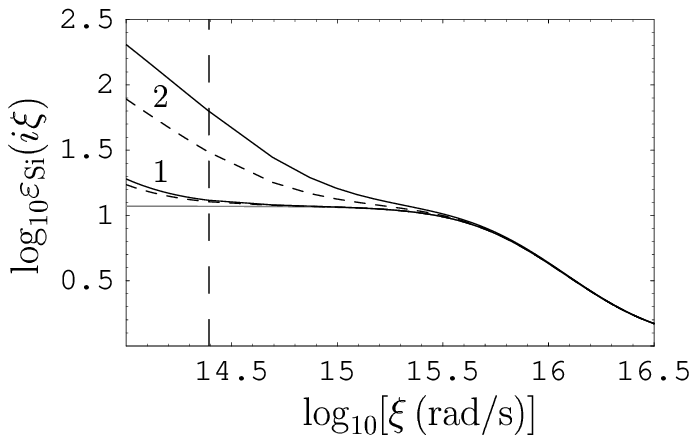}}
\vspace*{-12.cm}
\caption{
The dielectric permittivities of high-resistivity Si (the grey
line), of P-doped Si in the insulating
 and metallic states (dashed
and solid lines marked 1, respectively) and
 of S-doped Si in dielectric and metallic states (dashed
and solid lines marked 2, respectively) are plotted along the
imaginary frequency axis. The first Matsubara frequency at
300\,K is indicated by the vertical long-dashed line.
}
\end{figure*}
%%%%%%%%%%%%%%%

The dielectric
permittivity of doped Si along the imaginary frequency axis is
given by \cite{55Si}
\begin{equation}
\varepsilon_{\rm Si:d}(i\xi)=\varepsilon_{\rm Si}(i\xi)+
\frac{\omega_{p,\rm Si}^2}{\xi(\xi+\gamma_{\rm Si})},
\label{eq8}
\end{equation}
\noindent
where $\gamma_{\rm Si}$ is the relaxation parameter and the plasma
frequency is expressed in terms of the effective mass of an
electron $m^{\ast}$ and the concentration of free charge
carriers $n_{\rm fc}^{\rm Si}$ by the equation
\begin{equation}
\omega_{p,\rm Si}=
\frac{e\sqrt{n_{\rm fc}^{\rm Si}}}{\sqrt{\epsilon_0m^{\ast}}}.
\label{eq9}
\end{equation}
\noindent
Here, $\epsilon_0$ is the permittivity of vacuum and $e$ is the
electron charge.

Now we consider two S-doped Si plates with
$n_{\rm fc}^a=1.8\times 10^{20}\,\mbox{cm}^{-3}<n_{\rm fc;cr}^{\rm Si:S}$
and
$n_{\rm fc}^b=4.3\times 10^{20}\,\mbox{cm}^{-3}>n_{\rm fc;cr}^{\rm Si:S}$.
These are $n$-type semiconductors with $m^{\ast}=0.26m_e$
where $m_e$ is the mass of an electron.
{}From (\ref{eq9}) one obtains
$\omega_{p,\rm Si:S}^{(a)}=1.32\times 10^{15}\,$rad/s and
$\omega_{p,\rm Si:S}^{(b)}=2.32\times 10^{15}\,$rad/s
($\gamma_{\rm Si}=1.8\times 10^{13}\,$rad/s).
The respective dielectric permittivities (\ref{eq8}) along the
imaginary frequency axis are shown in figure~1 by the dashed and
solid lines marked 2. The first Matsubara frequency at $T=300\,$K
is indicated by the vertical long-dashed line.

For comparison purposes we also consider two P-doped Si plates
used in a previous experiment \cite{49}, but with doping
concentrations
$n_{\rm fc}^a=3.7\times 10^{18}\,\mbox{cm}^{-3}<n_{\rm fc;cr}^{\rm Si:P}$
and
$n_{\rm fc}^b=4\times 10^{18}\,\mbox{cm}^{-3}>n_{\rm fc;cr}^{\rm Si:P}$,
i.e., much closer to the critical value.
These are also $n$-type semiconductors characterized by the
same effective mass of an electron. The respective values of
the plasma frequency calculated using (\ref{eq9}) are
$\omega_{p,\rm Si:P}^{(a)}=2.81\times 10^{14}\,$rad/s and
$\omega_{p,\rm Si:P}^{(b)}=2.93\times 10^{14}\,$rad/s.
The respective dielectric permittivities
are shown in figure~1 by the dashed and
solid lines marked 1.

In all cases the material of the sphere (Au) is described by the
optical data for the complex index of refraction \cite{55Au}
extrapolated to lower frequencies by means of the Drude model
with $\omega_{p,\rm Au}=9.0\,$eV, $\gamma_{\rm Au}=0.035\,$eV.
It was recently shown \cite{57} that this extrapolation is in
excellent agreement with the optical data measured over a wide
frequency region.

\subsection{Force between an Au sphere and Si plates}

Now we are in a position to calculate the Casimir forces between
a sphere and each of the two different
Si plates one of which is in metallic
and  the other in an insulating state.
Keeping in mind that technologically there is no problem to
prepare Si plates of sufficient thickness, the quantity $d$ in
(\ref{eq4}) can be put equal to infinity so that the reflection
coefficients (\ref{eq5}) on a Si plate take the same form as
the coefficients (\ref{eq3}) on Au (the role of finite thickness
of a semiconductor film is considered in Sec.~4).

%%%%%%%%%%%%%%%
\begin{figure*}[h]
\vspace*{-13.cm}
\centerline{\includegraphics{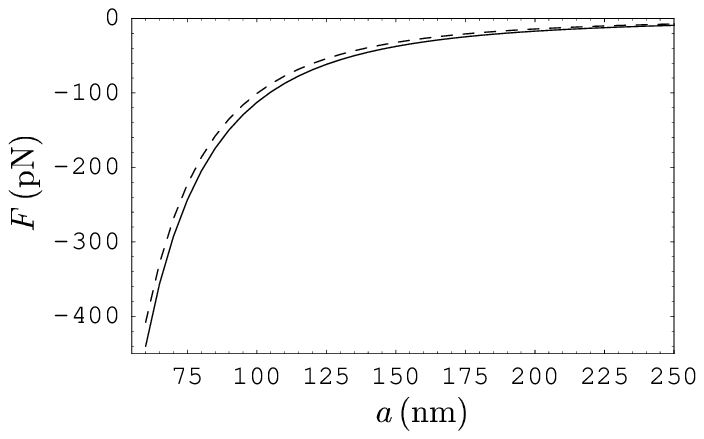}}
\vspace*{-12.cm}
\caption{
The Casimir forces between an Au-coated sphere and a plate made of
S-doped Si in metallic state (solid line) and in dielectric state
(dashed line) are shown as functions of separation when free charge
carriers in the insulating state are omitted.
}
\end{figure*}
%%%%%%%%%%%%%%%
The Casimir forces acting between an Au sphere of
$R=101.2\,\mu$m (such a sphere was used in experiments with
ITO films \cite{26,27,43}) and each of the two S-doped Si plates
was calculated by (\ref{eq2}). The force values in the case
of metallic-type Si decsribed by the dielectric permittivity
$\varepsilon_{\rm Si:S}^{(b)}$ (the solid line marked 2 in
figure~1) are shown by the solid line in figure~2 as a function
of separation.
The values of the Casimir force for a plate made of insulating-type
Si decribed by the permittivity $\varepsilon_{\rm Si}$ with free
charge carriers omitted (the grey line in figure~1) are shown by
the dashed line. This line is obtained similar in spirit to the
theoretical interpretation of experiments
\cite{26,27,37,38,39,40,43} discussed in Sec.~1.
As can be seen in figure~2, there is significant relative
deviation between the solid and dashed lines that can be
observed experimentally with present experimental
precision. Thus, the relative difference of sphere-plate Casimir
forces with metallic and dielectric plates,
$|F_{\rm met}-F_{\rm diel}|/|F_{\rm diel}|$, is equal to
7.2\%, 10.8\%, 15.0\%, 18.7\%, and 21.8\% at separations equal
to 60, 100, 150, 200, and 250\,nm, respectively.
At separations of 450 and 500\,nm this relative difference
reaches 30.2\% and 31.5\%, respectively.

This should be compared with the relative force differences
for a sphere-plate interaction when the two P-doped Si plates
with the doping concentrations $n_{\rm fc}^a$ and
$n_{\rm fc}^b$ indicated in
this section are used (see the solid line marked 1 and the grey
line in figure~1). At separation of
60, 100, 150, 200, and 250\,nm the relative force differences
are equal to 1.1\%, 1.6\%, 2.2\%, 2.8\%, and 3.4\%, respectively,
when the same calculation method is used (i.e., the contribution
of free charge carriers is omitted in the dielectric state).
At separations of 450 and 500\,nm the relative force difference
in the case of P-doped Si reaches only 5.7\% and 6.2\%,
respectively. Thus, the replacement of P-doped Si with
S-doped Si suggests considerable opportunities in the
experimental observation of the effect of insulator-metal
transition on the Casimir force in doped semiconductors.

To finally determine the feasibility of the proposed experiment
with the plates made of S-doped Si, we compare the absolute
changes of the Casimir force due to an insulator-metal
transition, $|F_{\rm met}-F_{\rm diel}|$, with the
total error in the force difference $2\Delta_F$.
Here $\Delta_F$ is the total experimental error in the measured
Casimir forces determined at a 67\% confidence level
\cite{26,27,43}.
In figure~3 we plot the quantity $|F_{\rm met}-F_{\rm diel}|$
calculated with omitted contribution of free charge carriers
in the dielectric state as a function of separation (solid line).
In the same figure the total error in the
force difference is
indicated by the grey line. As can be seen in figure 3, at
separations
$a=60$, 100, 150, 200, and 250\,nm we obtain
$|F_{\rm met}-F_{\rm diel}|=29.5$, 11.0, 4.9, 2.7, and 1.65\,pN,
respectively, whereas the respective total errors in the force
difference in the recently performed experiment with two ITP plates
\cite{26,27} (the grey line)
are equal to $2\Delta_F=3.0$, 2.0, 1.8, 1.75, and
1.6\,pN. This means that at separations below 250\,nm the
predicted changes in the Casimir force due to insulator-metal
transition in S-doped Si can be reliably detected by using
existing experimental setup.
%%%%%%%%%%%%%%%
\begin{figure*}[h]
\vspace*{-13.cm}
\centerline{\includegraphics{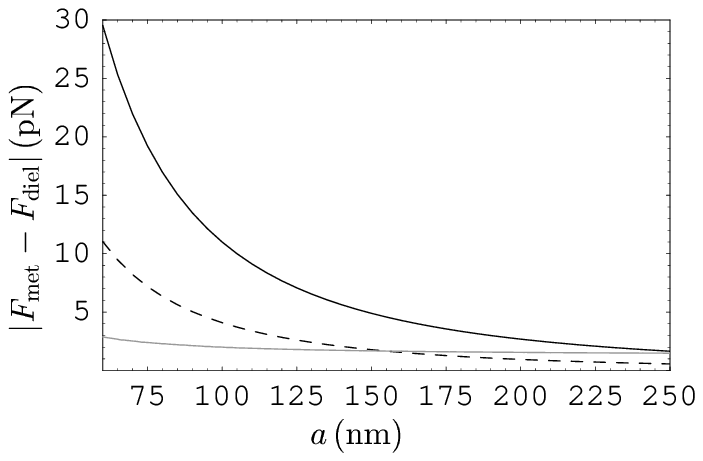}}
\vspace*{-12.cm}
\caption{
Differences in the Casimir force between an Au-coated sphere and two plates made of
S-doped Si in the metallic and insulating states, calculated with free charge
carriers in insulating state omitted (solid line) and included
(dashed line), are shown as functions of separation.
The total error in the force difference is shown by the grey line.
}
\end{figure*}
%%%%%%%%%%%%%%%

Note that if the exact value of the critical concentration
$n_{\rm fc;cr}^{\rm Si:S}$ is known we could choose the
concentrations $n_{\rm fc}^a$ and
$n_{\rm fc}^b$ in the insulating and metallic
plates, respectively, in such a way that the dielectric
permittivities
$\varepsilon_{\rm Si:S}^{(a)}$ and
$\varepsilon_{\rm Si:S}^{(b)}$
were almost coincident. In this case nonzero change in the magnitude
of the Casimir force due to insulator-metal transition occurs
only in the case that the charge carriers in the insulating state
are omitted. In our case, however, the proposal is based on the
findings of \cite{52} that the sample with charge carriers
concentration $n_{\rm fc}=1.8\times 10^{20}\,\mbox{cm}^{-3}$ is in
an insulating state, whereas the sample with
$n_{\rm fc}=4.3\times 10^{20}\,\mbox{cm}^{-3}$ is in the metallic state.
As a result, the respective dielectric permittivities are
noticeably different (see the dashed and solid lines marked 2
in figure~1).

Because of this, it is of much interest to compare the above
results obtained with the contribution of free charge
carriers of the dielectric plate omitted with the results computed
by the immediate application of the Lifshitz theory with no
additional prescriptions (the dashed line in figure 3).
In this case for S-doped Si at separations 60, 100, and 150\,nm
one obtains $|F_{\rm met}-F_{\rm diel}|=11.0$, 4.1, and
1.8\,pN, respectively, where metallic Si is described by the
permittivity $\varepsilon_{\rm Si:S}^{(b)}$ and dielectric Si
by $\varepsilon_{\rm Si:S}^{(a)}$ defined in (\ref{eq8})
(see the solid and dashed lines marked 2 in figure~1).
It can be seen that the immediate application of the standard
Lifshitz theory leads to much smaller force differences although
still detectable at separations below 150\,nm.
These differences are caused by the change of dielectric
permittivity, whereas the major contribution to the
differences calculated above was caused by the fact that
really existing free charge carriers in the dielectric state were
omitted.

For the plates made of P-doped Si the insulator-metal phase
transition does not lead to a detectable effect.
Thus, at separations $a=60$ and 100\,nm one obtains
$|F_{\rm met}-F_{\rm diel}|=4.45$ and 1.61\,pN, respectively,
if the contribution of free charge carriers in the dielectric
plate is omitted (i.e., the dielectric permittivities shown
by the solid line marked 1 and the grey line in figure 1 are
used in computations). It is hardly probable that this
effect can be detected and if yes at the shortest separations
only because
$2\Delta_F(100\,\mbox{nm})=2\,\mbox{pN}>1.61\,$pN.
As to the immediate application of the standard Lifshitz
theory (i.e., using the dielectric permittivities shown
by the solid and dashed lines marked 1 in figure 1),
at $a=60$ and 100\,nm it follows
$|F_{\rm met}-F_{\rm diel}|=0.08$ and 0.03\,pN, respectively,
which is not detectable in the foreseeable future.
Thus, the use of Si doped with S offers excellent
possibilities to directly check the predictions of the
Lifshitz theory for semiconductors which undergo
the insulator-metal transition. Dynamic measurements considered
in the next section suggest some additional opportunities
in this respect.

\subsection{Pressure between an Au plate and  Si plates}

Now we consider the possibilities to observe the effect of
insulator-metal transition in a doped semiconductor on the
Casimir pressure. As explained in Sec.~2, the Casimir pressure
between two parallel plates is obtained from the immediately
measured gradient of the Casimir force between a sphere and
a plate in the dynamic regime. For this purpose we first
compute the Casimir pressure (\ref{eq6}) between an Au plate
and Si plate using the dielectric permittivity of metallic Si
doped with S shown by the solid line marked 2 in figure~1.
The magnitude of the obtained pressure is shown by the solid
line in figure 4.
%%%%%%%%%%%%%%%
\begin{figure*}[h]
\vspace*{-13.cm}
\centerline{\includegraphics{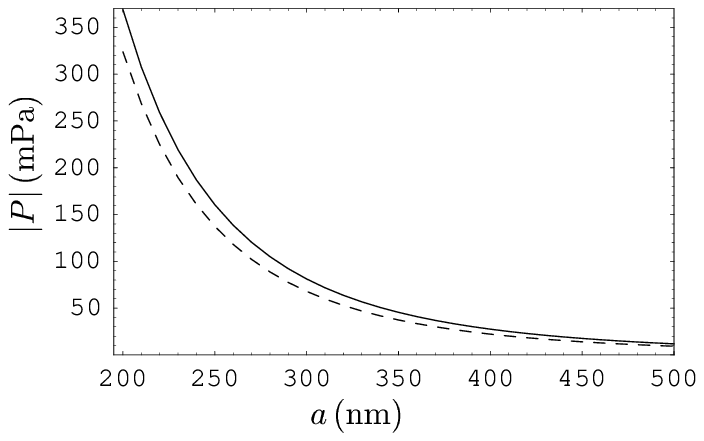}}
\vspace*{-12.cm}
\caption{
The magnitudes of the Casimir pressure between an Au plate and a plate made of
S-doped Si in metallic state (solid line) and in dielectric state
(dashed line) are shown as functions of separation when free charge
carriers in the insulating state are omitted.
}
\end{figure*}
%%%%%%%%%%%%%%%
For insulating Si with the contribution of charge carriers
 omitted, the dielectric permittivity along the imaginary
 frequency axis (\ref{eq7}) is shown by the grey line in
 figure~1. The computational results for the magnitude of
 the Casimir pressure as a function of separation are shown
 by the dashed line in figure 4.
 Here, larger separation distances, than in figure~2 are chosen
(as are usually used in dynamic experiments where the separation
distance between the sphere and the plate is varied harmonically
with time). As can be seen in figure 4, there are significant
relative changes in the Casimir pressure due to the insulator-metal
transition. Thus, at separations 200, 250, 300, and 350\,nm
the quantity $|P_{\rm met}-P_{\rm diel}|/|P_{\rm diel}|$
takes the values 13.9\%, 16.8\%, 19.5\%, and 21.8\%,
respectively. At $a=500\,$nm the relative change in the pressure
reaches 27.6\%. This is somewhat smaller than for the Casimir
force. However,  dynamic
experiments are more precise than static ones. Because of
this, the effect of the phase transition can be observed at
larger separations.

To see this, we calculate the absolute change in the Casimir
pressure, $|P_{\rm met}-P_{\rm diel}|$,  under the same
conditions as discussed above, i.e., omitting the contribution
of free charge carriers in dielectric Si. The values of
$|P_{\rm met}-P_{\rm diel}|$ as a function of separation are
shown in figure~5 by the solid line. As a result, at
separations $a=200$, 250, 300, 350, 400, and 450\,nm
the absolute change in the Casimir pressures takes the values
45.1, 23.1, 13.2, 8.2, 5.3, and 3.6\,mPa, respectively.
In the recently performed experiment on measuring the
Casimir pressure by means of a dynamic AFM operated in the
frequency modulation technique \cite{31b,32} the total
experimental error in the measured pressure
does not depend on separation and is equal to
$\Delta_P=1.9\,$mPa at a 67\% confidence level.
Thus, the total error in the pressure difference is
$2\Delta_P=3.8\,$mPa .
It is shown by the grey line in figure~5. By comparing
this error with the absolute changes in the Casimir pressure
listed above (the solid line in figure 5), we arrive at the
conclusion that with the existing setup the effect of
the insulator-metal transition on the Casimir pressure can be
observed over a separation region from 200\,nm to
approximately 440\,nm.
%%%%%%%%%%%%%%%
\begin{figure*}[h]
\vspace*{-13.cm}
\centerline{\includegraphics{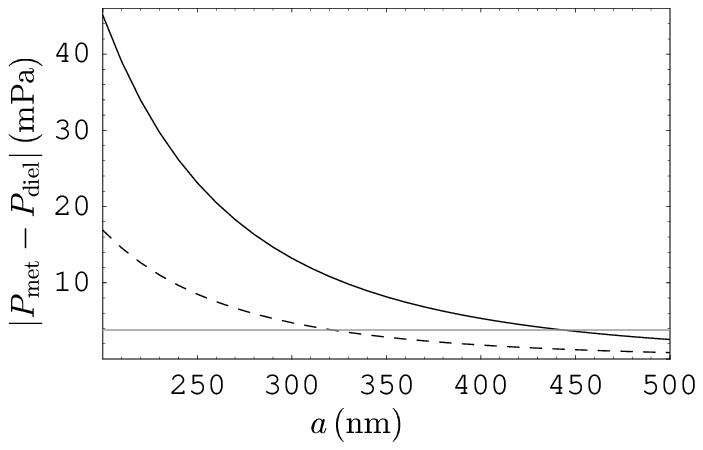}}
\vspace*{-12.cm}
\caption{
Differences in the Casimir pressure between an Au-coated sphere and two plates made of
S-doped Si in metallic and insulating states, calculated with free charge
carriers in the insulating state omitted (solid line) and included
(dashed line), are shown as functions of separation.
The total error in the pressure difference is shown by the grey line.
}
\end{figure*}
%%%%%%%%%%%%%%%

It is interesting to compare this prediction with the change in
the Casimir pressure following from the immediate application of
the Lifshitz theory with no omissions. In this case the dielectric
 permittivities shown by the solid and dashed lines marked 2 in
figure~1 are substituted in expression (\ref{eq6}) for the
Casimir pressure. The obtained pressure differences
$|P_{\rm met}-P_{\rm diel}|$ are plotted in figure~5 by the
dashed line. As a result, at
separations $a=200$, 250, 300, and 350\,nm the following changes
in the magnitude of the Casimir pressure due to dielectric-metal
transition are obtained:
$|P_{\rm met}-P_{\rm diel}|=16.9$, 8.5, 4.7, and 2.9\,mPa.
It can be seen that these changes resulting from real differences
in dielectric permittivities of the plates made of S-doped Si
are several times less than the changes computed above with the
contribution of free charge carriers in dielectric Si omitted.
However, from the comparison with the total error in the pressure
difference
$2\Delta_P=3.8\,$mPa (the grey line in figure 5), one can conclude
 that even this smaller effect can be observed by means of
 dynamic measurements at separations from 200 to approximately
 315\,nm. This makes the proposed experiment vitally important
 for choosing between competing theoretical approaches.

 Computations in this section are done for Si doped with S.
 Almost the same numerical results are obtained for Si plates
 doped with Se. As mentioned in Sec.~2, this semiconductor also
 has high critical doping concentration of order
 $10^{20}\,\mbox{cm}^{-3}$. Because of this, it is also
 prospective for use in experiments on the impact of
 Mott-Anderson phase transition on the Casimir effect.

 It is worth mentioning also that if metal of the sphere
 (Au) is described by the generalized plasma-like model
\cite{2,21}, which leads to theoretical predictions
consistent with the experimental data obtained for two
metallic bodies \cite{28,29,30,31,31b,32}, this does not
change anything in our conclusions or numerical results.

\section{Change of dispersion forces due to the insulator-metal
transition in B-doped diamond}

As discussed in Sec.~2, another prospective semiconductor material
 to investigate the influence of insulator-metal transition on the
 Casimir force, is B-doped diamond. Boron-doped diamond films
 are usually deposited on silica (SiO${}_2$) substrates. For our
 purposes it is sufficient to use simple analytic approximations
for the dielectric permittivity of both materials along the
imaginary frequency axis (the experimental samples should be
investigated by means of ellipsometry). Thus, for diamond with no
doping we get \cite{58}
\begin{equation}
\varepsilon_{\rm C}(i\xi)=1+
\frac{C_{\rm UV}}{1+\frac{\xi^2}{\omega_{\rm UV}^2}}+
\frac{C_{\rm IR}}{1+\frac{\xi^2}{\omega_{\rm IR}^2}},
\label{eq10}
\end{equation}
\noindent
where $C_{\rm UV}=4.642$, $C_{\rm IR}=0.02$,
$\omega_{\rm UV}=1.61\times 10^{16}\,$rad/s, and
$\omega_{\rm IR}=2.5\times 10^{14}\,$rad/s.
The respective static dielectric permittivity is equal to
$\varepsilon_{\rm C,0}=5.66$. In figure~6 the behavior of
$\varepsilon_{\rm C}$ is shown by the grey line.
%%%%%%%%%%%%%%%
\begin{figure*}[h]
\vspace*{-13.cm}
\centerline{\includegraphics{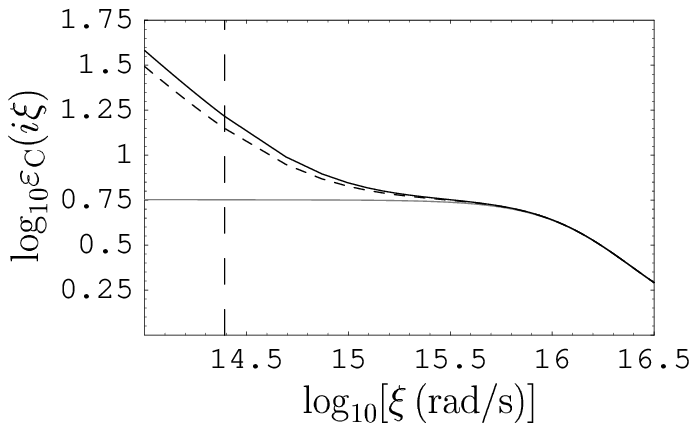}}
\vspace*{-12.cm}
\caption{
The dielectric permittivities of high-resistivity diamond (the grey
line) and of B-doped diamond in dielectric and metallic states (dashed
and solid lines, respectively) are plotted along the
imaginary frequency axis. The first Matsubara frequency at
300\,K is indicated by the vertical long-dashed line.
}
\end{figure*}
%%%%%%%%%%%%%%%

The dielectric permittivity of the B-doped diamond is represented
similar to (\ref{eq8})
\begin{equation}
\varepsilon_{\rm C:B}(i\xi)=\varepsilon_{\rm C}(i\xi)+
\frac{\omega_{p,\rm C}^2}{\xi(\xi+\gamma_{\rm C})},
\label{eq11}
\end{equation}
\noindent
where $\varepsilon_{\rm C}(i\xi)$ is defined in (\ref{eq10}).
We consider two B-doped diamond films with
$n_{\rm fc}^a=3.95\times 10^{20}\,\mbox{cm}^{-3}<n_{\rm fc;cr}^{\rm C:B}$
and
$n_{\rm fc}^b=5.05\times 10^{20}\,\mbox{cm}^{-3}>n_{\rm fc;cr}^{\rm C:B}$
(see Sec.~2). The respective values of the plasma frequency
calculated using (\ref{eq9}) rewritten for C instead of Si are
$\omega_{p,\rm C:B}^{(a)}=1.30\times 10^{15}\,$rad/s and
$\omega_{p,\rm C:B}^{(b)}=1.47\times 10^{15}\,$rad/s.
Here, the effective mass $m^{\ast}=0.74m_e$ has been used \cite{54}.
 The relaxation parameter in (\ref{eq11}) is equal to
$\gamma_{\rm C}=5.65\times 10^{14}\,$rad/s \cite{59}.
The dielectric permittivities of metallic and dielectric
B-doped diamond films are shown in figure~6 by the solid and
dashed lines, respectively.
The first Matsubara frequency at
300\,K is indicated by the vertical long-dashed line.

The parametrization of the dielectric permittivity of SiO${}_2$
is given by  \cite{58}
\begin{equation}
\varepsilon_{{\rm SiO}_2}(i\xi)=1+
\frac{C_{\rm UV}}{1+\frac{\xi^2}{\omega_{\rm UV}^2}}+
\sum_{i=1}^{3}\frac{C_{\rm IR}^{(i)}}{1+\frac{\xi^2}{\omega_{{\rm IR},i}^2}},
\label{eq12}
\end{equation}
\noindent
where $C_{\rm UV}=1.098$, $\omega_{\rm UV}=2.034\times 10^{16}\,$rad/s,
$C_{\rm IR}^{(1)}=0.829$, $C_{\rm IR}^{(2)}=0.095$,
$C_{\rm IR}^{(3)}=0.798$,
$\omega_{\rm IR,1}=0.867\times 10^{14}\,$rad/s,
$\omega_{\rm IR,2}=1.508\times 10^{14}\,$rad/s, and
$\omega_{\rm IR,3}=2.026\times 10^{14}\,$rad/s.
For the static dielectric permittivity of SiO${}_2$ one obtains
$\varepsilon_{{\rm SiO}_2,0}=3.92$.
In computations below we made sure that the thickness $d$ of a
B-doped diamond film only slightly influences the values of the
Casimir force. The material of the sphere (Au) is described in the
same way as in Sec.~3 (i.e., by the tabulated optical data
extrapolated to lower frequencies using the Drude model).

\subsection{Force between an Au sphere and B-doped
diamond films}

Now we calculate the Casimir force acting between an Au sphere
of $R=101.2\,\mu$m radius and each of the two B-doped diamond
films deposited on a SiO${}_2$ substrate using (\ref{eq2}).
In the case of a metallic diamond film the Casimir force as a
function of separation is shown by the solid line in figure~7.
It was computed with the dielectric permittivity
$\varepsilon_{\rm C:B}^{(b)}$ (the solid line in figure~6).
For the dielectric diamond film the computational results
are shown in figure~7 by the dashed line. These results were
obtained with the permittivity $\varepsilon_{\rm C}$
(the grey line in figure~6) which does not take into account
the free charge carriers. There are significant relative
deviations between the two lines in figure~7. Thus, at
separations $a=60$, 100, 150, 200, 300, and 400\,nm the
quantity $|F_{\rm met}-F_{\rm diel}|/|F_{\rm diel}|$ is
equal to 8.4\%, 12.1\%, 16.3\%, 20.0\%, 26.2\%, and 31.1\%,
respectively. At $a=500\,$nm this quantity reaches 35.0\%.
%%%%%%%%%%%%%%%
\begin{figure*}[h]
\vspace*{-13.cm}
\centerline{\includegraphics{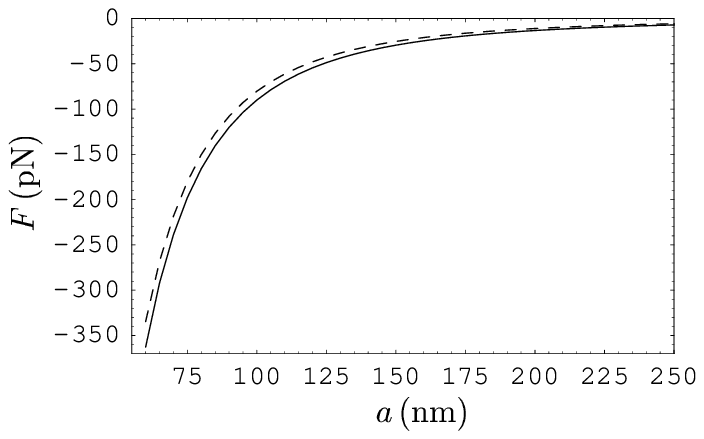}}
\vspace*{-12.cm}
\caption{
The Casimir forces between an Au-coated sphere and a SiO${}_2$
substrate coated with a thick
B-doped diamond film in the metallic state (solid line) and in
the insulating state
(dashed line) are shown as functions of separation when free charge
carriers in dielectric state are omitted.
}
\end{figure*}
%%%%%%%%%%%%%%%

The above computations were performed for an infinitely thick film
($d=\infty$). To determine the role of finite thickness of the
diamond film, we calculated the same relative differences in the
Casimir force for a film of $d=100\,$nm thickness deposited on a
silica substrate. Diamond films of  about 100\,nm thickness
are in fact of the minimum thickness used in technological applications.
At separations $a=60$ and 100\,nm the relative difference in the
Casimir force is the same as for an infinitely thick film.
Minor changes between these two cases arise only with the increase
of separation distance. Thus, for a film
of $d=100\,$nm thickness at
separations $a=150$,  200, 300, and 400\,nm  the
quantity $|F_{\rm met}-F_{\rm diel}|/|F_{\rm diel}|$ is
equal to 16.4\%, 20.2\%, 26.8\%, and 32.3\%,
respectively. At $a=500\,$nm  for a film
of $d=100\,$nm thickness the relative difference is equal to 36.8\%.
Comparing with the above results obtained for an infinitely thick
semispace made of B-doped diamond,
one can conclude that the B-doped diamond film of 100\,nm
 thickness leads to almost the same effect as a semispace.

To determine the feasibility of the proposed experiment with
B-doped diamond films, we present the absolute changes in the
Casimir force $|F_{\rm met}-F_{\rm diel}|$ due to insulator-metal
transition. This quantity as a function of separation is plotted
in figure~8 (solid line). The total error in the force difference
(the same as in figure~3) is plotted by the grey line.
For a B-doped diamond semispaces at
separations $a=60$, 100, 150, 200, and 250\,nm the calculated
absolute changes are equal to 28.1, 9.7, 4.1, 2.2, and 1.4\,pN,
respectively. For a diamond film of 100\,nm thickness the
respective force changes are 27.7, 9.5, 4.0, 2.1, and 1.3\,pN.
Thus, the predicted effect can be observed using the
existing setup within the separation region from 60 to
approximately 225\,nm.
%%%%%%%%%%%%%%%
\begin{figure*}[h]
\vspace*{-13.cm}
\centerline{\includegraphics{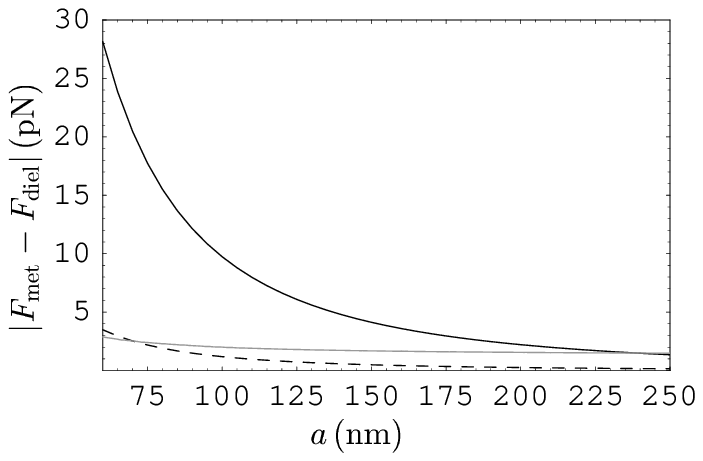}}
\vspace*{-12.cm}
\caption{
Differences in the Casimir force between an Au-coated sphere and two plates made of
a SiO${}_2$
substrate coated with thick
B-doped diamond films in metallic and insulating states, calculated with free charge
carriers in the insulating state omitted (solid line) and included
(dashed line), are shown as functions of separation.
The total error in the force difference is shown by the grey line.
}
\end{figure*}
%%%%%%%%%%%%%%%

We have also computed the change in the Casimir force
$|F_{\rm met}-F_{\rm diel}|$ predicted by the standard Lifshitz
theory, i.e., when $F_{\rm met}$ is obtained using the dielectric
permittivity $\varepsilon_{\rm C:B}^{(b)}$ (the solid line in
figure~6) and $F_{\rm diel}$ is obtained using the dielectric
permittivity $\varepsilon_{\rm C:B}^{(a)}$ (the dashed line in
figure~6). This prediction of the standard theory is shown by the
dashed line in figure~8. In this case at separations $a=60$ and
100\,nm we obtain $|F_{\rm met}-F_{\rm diel}|=3.5$ and 1.2\,pN,
respectively. This means that the predictions of the standard
Lifshitz theory caused by the change of dielectric permittivity
exceed the total experimental error only at the shortest separations
below about 70\,nm.

\subsection{Pressure between an Au plate and a
B-doped diamond plate}

Here we discuss the possibilities to observe the change of the
Casimir pressure in the insulator-metal transition in diamond films
using an AFM operated in the dynamic regime. For the dielectric
film computations were done with the contribution of free charge
carriers omitted (the grey line in figure~6). The obtained
Casimir pressure versus separation is shown by the dashed line
in figure~9. For a metallic film the dielectric
permittivity $\varepsilon_{\rm C:B}^{(b)}(i\xi)$
was used (the solid line in
figure~6). In this case the Casimir pressure as a function of
separation is presented by the solid line in figure~9.
The relative deviations in the Casimir pressure
$|P_{\rm met}-P_{\rm diel}|/|P_{\rm diel}|$ vary from
15.2\% and 17.9\% at separations 200 and 250\,nm, respectively,
to 29\% at $a=500\,$nm.
%%%%%%%%%%%%%%%
\begin{figure*}[h]
\vspace*{-13.cm}
\centerline{\includegraphics{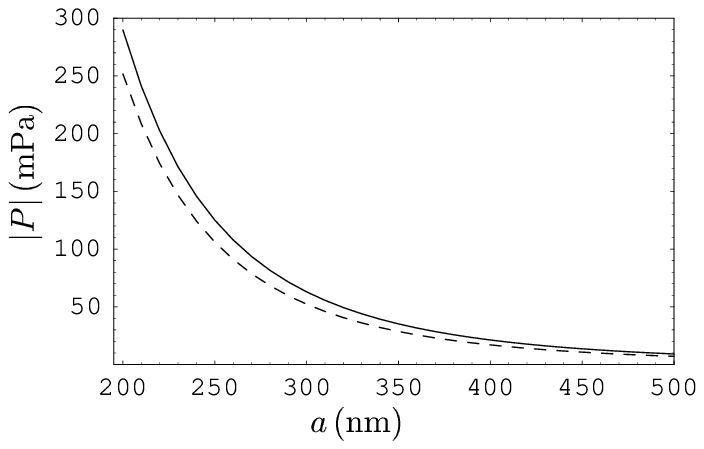}}
\vspace*{-12.cm}
\caption{
The magnitudes of the Casimir pressure between an Au plate and a SiO${}_2$
substrate coated with thick
B-doped diamond film in the metallic state (solid line) and in
the insulating state
(dashed line) are shown as functions of separation when free charge
carriers in the insulating state are omitted.
}
\end{figure*}
%%%%%%%%%%%%%%%

All the above computations were performed at separations
$a\geq 60\,$nm where the effect of the phase transition is more
pronounced. Qualitatively the same effect, however, takes
place at separations of a few nanometers, i.e., in the region
of nonretarded van der Waals force. To demonstrate this,
we have computed the quantity
$|P_{\rm met}-P_{\rm diel}|/|P_{\rm diel}|$
at $a=3$, 5, and 10\,nm and found that it is equal to 1.9\%,
2.0\%, and 2.4\%, respectively.

We have also calculated the absolute change in the Casimir pressure
$|P_{\rm met}-P_{\rm diel}|$ with the contribution of free charge
 carriers of the dielectric plate omitted. The computational results
are shown by the solid line in figure~10. In the same figure the
total experimental error for the difference in Casimir
pressures determined at a 67\% confidence level is shown by the
grey line (the same as in figure~5). As can be seen in this
figure, the predicted effect can be easily observed using
existing laboratory setup over a wide separation region from
200 to 410\,nm. Numerically the calculated change in the Casimir
pressure is equal to 38.1, 19.0, 10.7, 6.6, and 4.3\,mPa at
separations $a=200$, 250, 300, 350, and 400\,nm, respectively.
These should be compared with the total error in the pressure
difference equal to 3.8\,mPa which does not depend on separation in
the case of the dynamic AFM \cite{32}.
%%%%%%%%%%%%%%%
\begin{figure*}[h]
\vspace*{-13.cm}
\centerline{\includegraphics{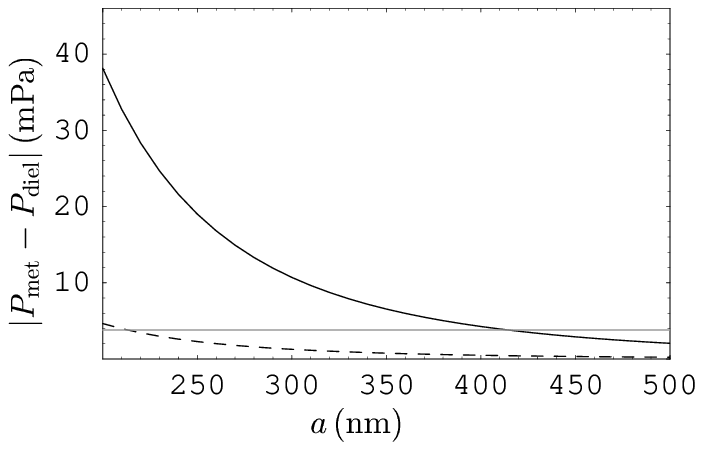}}
\vspace*{-12.cm}
\caption{
Differences in the Casimir pressure between an Au-coated sphere and two plates made of
a SiO${}_2$
substrate coated with thick
B-doped diamond films in the metallic and insulating states, calculated with
free charge
carriers in the insulating state omitted (solid line) and included
(dashed line), are shown as functions of separation.
The total error in the force difference is shown by the grey line.
}
\end{figure*}
%%%%%%%%%%%%%%%

The complete application of the Lifshitz theory, i.e., the use of
 dielectric permittivities $\varepsilon_{\rm C:B}^{(a)}$ and
$\varepsilon_{\rm C:B}^{(b)}$ for dielectric and metallic diamond
films, respectively, leads to much smaller changes in the Casimir
pressure. They are shown by the dashed line in figure~10.
Specifically, at separations $a=200$ and 250\,nm the difference
$|P_{\rm met}-P_{\rm diel}|$ computed according to the standard
Lifshitz theory is equal to 4.6 and 2.3\,mPa, respectively.
As can be seen in figure~10, this effect is observable only at
the shortest separations ranging from 200 to about 210\,nm.

The above computations show that the use of B-doped diamond
suggests similar opportunities as the use of Si doped with S
or Se with respect to observation of the change in the Casimir
force due to the insulator-metal transition. Because of this,
choice between the three materials should be done for reasons
of convenience in preparation of the test bodies.

\section{Conclusions and discussion}

In the foregoing we have considered the possible influence of the
Mott-Anderson insulator-metal transition  in doped semiconductors
on the Casimir force and Casimir pressure.
This subject was suggested by the unexpected results of several
experiments which demonstrated that to  bring the measurement
data in agreement with computations using the Lifshitz theory
one should omit the contribution of free charge carriers in
dielectric test bodies. In all the experiments performed to date
the dielectric permittivities of a semiconductor plate in the
metallic and dielectric states were significantly different.
In the proposed experiment using doped semiconductors with
high critical concentration of charge carriers, we put
forward the possibility to make this difference as small as
possible and simultaneously increase the effect from omission
of charge carriers in the dielectric state up to the values
easily detectable using existing laboratory setups.

We have suggested three different semiconductor materials
suitable for use in the proposed esperiment: Si doped with S or
Se and diamond doped with B. All these materials offer very
prospective opportunities for the observation of the effect
under discussion. Thus, for a S-doped Si the calculated
difference between the Casimir forces in metallic and dielectric
states at separations 60, 100, and 150\,nm exceeds the total
experimental error by a factor of 9.8, 5.5, and 2.7,
respectively. For the Casimir pressures at separations 200,
250, and 300\,nm the similar difference in the
Casimir pressures exceeds the total experimental error by a
factor of 11.9, 6.1, and 3.5, respectively.
 According to the standard Lifshitz theory
the differences between the Casimir forces and pressures in
the insulator-metal transition are entirely determined by the
differences in the measured dielectric properties of the
test bodies. Because of this, the standard Lifshitz theory
predicts much smaller changes in the Casimir force and
pressure due to insulator-metal transition in S-doped Si.
Specifically, for the same concentrations of charge
carriers in the dielectric and metallic states, as are used
above, the difference in the Casimir forces at separations of
60 and 100\,nm exceeds the total experimental error by a
factor of 3.7 and 2, respectively. At $a=150\,$nm this
difference is equal to the total error.
The pressure differences computed using the standard
Lifshitz theory at $a=200$, 250, and 300\,nm exceed the
total experimental error by a factor of 4.4, 2.2, and 1.2,
respectively. At $a=350\,$nm the total error of the
difference in the Casimir pressures exceeds
the magnitude of the predicted difference.
Thus, theoretical predictions of the two approaches
in the experiment proposed differ
significantly and their validity can be tested with
confidence by comparison with the measurement data.

Finally, we would like to stress that the standard
Lifshitz theory does not allow
modifications of the Casimir
force without change of dielectric permittivity of
the test bodies over  a wide frequency region.
For the case, however, where the contribution of free
charge carriers in the dielectric permittivity of the
insulating materials should be omitted (as suggested
by several experiments discussed in Sec.~1), it
becomes possible to achieve significant
modifications in the force magnitude with only
negligibly small changes in the dielectric permittivity.
This happens due to the Mott-Anderson
insulator-metal transition in doped semiconductors when the
permittivities of insulating and metallic plates
differ only slightly. The proposed experiment
seems capable of providing definitive answer to
the question whether it is possible to modify
dispersion forces without change of dielectric
permittivity.

\ack{This work was supported by the  NSF Grant
No.~PHY0970161, DOE Grant
No.~DEF010204ER46131
and DARPA Grant under Contract
No.~S-000354 (U.M.).}
%\hspace*{-.5cm}

%%%%%%%%%%%%%%%%%%%%%%%%%%%%%%%%%%%

%%%%%

\section*{References}
\begin{thebibliography}{10}
\bibitem{1}
Parsegian V A 2005
{\it Van der Waals forces: A Handbook for Biologists,
Chemists, Engineers, and Physicists}
(Cambridge: Cambridge University Press)
\bibitem{2}
Bordag M, Klimchitskaya G L, Mohideen U and
Mostepanenko V M 2009
{\it Advances in the Casimir Effect}
(Oxford: Oxford University Press)
\bibitem{3}
Buks E and Roukes M L 2001
{\it Phys. Rev.} B {\bf 63} 033402
\bibitem{4}
Schr\"{o}der E and Hyldgaard P 2003
{\it Mater. Sci. Engineer.} C {\bf 23} 721
\bibitem{5}
Schr\"{o}der E and Hyldgaard P 2003
{\it Surf. Sci.} {\bf 532-535} 880
\bibitem{6}
Kara A and Rahman T S 2005
{\it Surf. Sci. Rep.} {\bf 56} 159
\bibitem {7}
Blagov E V, Klimchitskaya G L and Mostepanenko V M
2005
{\it Phys. Rev.} B {\bf 71} 235401
\bibitem {8}
Blagov E V, Klimchitskaya G L and Mostepanenko V M
2007
{\it Phys. Rev.} B {\bf 75} 235413
\bibitem{8a}
Bimonte G, Calloni E, Esposito G, Milano L and
Rosa L 2005
{\it Phys. Rev. Lett.} {\bf 94} 180402
\bibitem{8b}
Bimonte G, Calloni E, Esposito G  and Rosa L 2005
{\it Nucl. Phys.} B {\bf 726} 441
\bibitem {9}
Bezerra V B, Klimchitskaya G L, Mostepanenko V M
and Romero C 2010
{\it Phys. Rev.} D {\bf 81} 055003
\bibitem {10}
Bezerra V B, Klimchitskaya G L, Mostepanenko V M
and Romero C 2011
{\it Phys. Rev.} D {\bf 83} 075004
\bibitem {10a}
 Mostepanenko V M, Bezerra V B, Klimchitskaya G L
and Romero C 2011
{\it Int. J. Mod. Phys.} A {\bf 27} 1260015
\bibitem {11}
London F 1930
{\it Z. Phys.} {\bf 63} 245
\bibitem {12}
Casimir H B G 1948
{\it Proc. K. Ned. Akad. Wet.} B
{\bf 51} 793
\bibitem {13}
Langreth D C, Dion M, Rydberg H, Schr\"{o}der E,
Hyldgaard P and Lundqvist B I 2005
{\it Int. J. Quant. Chem.} {\bf 101} 599
\bibitem{14}
Lifshitz E M 1956
{\it Zh. Eksp. Teor. Fiz.} {\bf 29} 94
(1955 {\it Sov. Phys. JETP}  {\bf 2} 73)
\bibitem{15}
Lifshitz E M and Pitaevskii L P  1984
{\it Statistical Physics}, Pt II (Oxford: Pergamon Press)
\bibitem{16}
Emig T, Jaffe R L, Kardar M and
Scardicchio A 2006
{\it Phys. Rev. Lett.} {\bf 96} 080403
\bibitem{17}
Emig T, Graham N, Jaffe R L and Kardar M  2007
{\it Phys. Rev. Lett.} {\bf 99} 017403
\bibitem{18}
Kenneth O and Klich I 2008
{\it Phys. Rev.} B {\bf 78} 014103
\bibitem{19}
Bimonte G, Emig T, Jaffe R L and Kardar M 2012
{\it Europhys. Lett.} {\bf 97} 50001
\bibitem{20}
Bimonte G, Emig T and Kardar M 2012
{\it Appl. Phys. Lett.} {\bf 100}  074110
\bibitem{21}
Klimchitskaya G L, Mohideen U and
Mostepanenko V M  2009
{\it Rev. Mod. Phys.} {\bf 81} 1827
\bibitem{22}
Klimchitskaya G L, Mohideen U and
Mostepanenko V M  2011
{\it Int J. Mod. Phys.} B {\bf 25} 171
\bibitem{23}
Rodriguez A W, Capasso F and Johnson S G 2011
{\it Nature Photon.} {\bf 5} 211
\bibitem{24}
Chiu H-C, Klimchitskaya G L, Marachevsky V N,
Mostepanenko V M and Mohideen U 2009
{\it Phys. Rev.} B  {\bf 80} 121402(R)
\bibitem{25}
Chiu H-C, Klimchitskaya G L, Marachevsky V N,
Mostepanenko V M and Mohideen U 2010
{\it Phys. Rev.} B  {\bf 81} 115417
\bibitem{26}
Chang C-C, Banishev A A,
Klimchitskaya G L, Mostepanenko V M and Mohideen U 2011
{\it Phys. Rev. Lett.}  {\bf 107} 090403
\bibitem{27}
Banishev A A, Chang C-C,  Castillo-Garza R,
Klimchitskaya G L, Mostepanenko V M and Mohideen U 2012
{\it Phys. Rev.} B {\bf 85} 045436
\bibitem{27a}
Banishev A A, Chang C-C,
Klimchitskaya G L, Mostepanenko V M and Mohideen U 2012
{\it Phys. Rev.} B {\bf 85} 195422
\bibitem{28}
Decca R S, Fischbach E, Klimchitskaya G L,
 Krause D E, L\'opez D and Mostepanenko V M 2003
{\it Phys. Rev.} D {\bf 68} 116003
\bibitem{29}
Decca R S, L\'opez D, Fischbach E, Klimchitskaya G L,
 Krause D E and Mostepanenko V M 2005
 {\it  Ann. Phys. NY } {\bf 318} 37
\bibitem{30}
Decca R S, L\'opez D, Fischbach E, Klimchitskaya G L,
 Krause D E and Mostepanenko V M 2007
 {\it  Phys. Rev} D {\bf 75} 077101
\bibitem{31}
Decca R S, L\'opez D, Fischbach E, Klimchitskaya G L,
 Krause D E and Mostepanenko V M 2007
{\it Eur. Phys. J.} C {\bf 51} 963
\bibitem{31a}
Bostr\"{o}m M and Sernelius B E 2000
{\it Phys. Rev. Lett.} {\bf 84} 4757
\bibitem{31b}
Chang C-C, Banishev A A, Castillo-Garza R,
Klimchitskaya G L, Mostepanenko V M and Mohideen U 2012
{\it Phys. Rev.} B {\bf 85} 165443
\bibitem{32}
Chang C-C, Banishev A A, Castillo-Garza R,
Klimchitskaya G L, Mostepanenko V M and Mohideen U 2012
{\it Int. J. Mod. Phys.: Conf. Ser.} {\bf 14} 270
\bibitem{33}
Masuda M and Sasaki M 2009
{\it Phys. Rev. Lett.} {\bf 102} 171101
\bibitem{34}
 Sushkov A O, Kim W J, Dalvit D A R
and Lamoreaux S K 2011
{\it Nature Phys.} {\bf 7} 230
\bibitem {35}
Bezerra V B, Klimchitskaya G L, Mohideen U,
Mostepanenko V M and Romero C 2011
{\it Phys. Rev.} B {\bf 83} 075417
\bibitem{36}
Klimchitskaya G L, Bordag M,  Fischbach E,
 Krause D E and Mostepanenko V M 2011
{\it Int. J. Mod. Phys.} A {\bf  26} 3918
\bibitem{36a}
Klimchitskaya G L, Bordag M
and Mostepanenko V M 2012
{\it Int. J. Mod. Phys.} A {\bf  27} 1260012
\bibitem{36b}
Garcia-Sanches D, Fong K Y, Bhaskaran H, Lamoreaux S and
Tang H X 2012
{\it Phys. Rev. Lett.} {\bf 109} 027202
\bibitem{36c}
 Bordag M, Klimchitskaya G L
and Mostepanenko V M 2012
arXiv:1208.1757v1
\bibitem{37}
Chen F,  Klimchitskaya G L,
Mos\-te\-pa\-nen\-ko V M and Mohideen U 2007
{\it Optics Express} {\bf 15} 4823
\bibitem{38}
Chen F,  Klimchitskaya G L,
Mos\-te\-pa\-nen\-ko V M and Mohideen U 2007
{\it Phys. Rev.} B {\bf 76} 035338
\bibitem{39}
Obrecht J M, Wild R J, Antezza M, Pitaevskii L P,
Stringari S and Cornell E A 2007
{\it Phys. Rev. Lett.} {\bf 98} 063201
\bibitem{40}
Klimchitskaya G L and Mostepanenko V M 2008
{\it J. Phys. A: Math. Theor.} {\bf 41} 312002
\bibitem{43}
Banishev A A, Chang C-C,  Castillo-Garza R,
Klimchitskaya G L, Mostepanenko V M and Mohideen U 2012
{\it Int. J. Mod. Phys.} A {\bf 27} 1260001
\bibitem {44}
Bezerra V B, Klimchitskaya G L and
Mostepanenko V M 2002
{\it Phys. Rev.} A {\bf 66} 062112
\bibitem{45}
de~Man S, Heeck K, Wijngaarden R J and Iannuzzi D 2009
{\it Phys. Rev. Lett.} {\bf 103} 040402
\bibitem{46}
de~Man S, Heeck K and Iannuzzi D 2010
{\it Phys. Rev.} A {\bf 82} 062512
\bibitem{47}
 Geyer B, Klimchitskaya G L and
Mostepanenko V M
2005 {\it Phys. Rev.} D {\bf 72} 085009
\bibitem{45a}
Mostepanenko V M and Klimchitskaya G L 2010
{\it Int. J. Mod. Phys.} A {\bf 25} 2302
\bibitem{47a}
Shklovskii B I and Efros A L 1984
{\it Electronic Properties of Doped Semiconductors.
Solid State Series}, v.45 (Berlin: Springer)
\bibitem{48}
Toma\v{s} M S 2002
{\it Phys. Rev.} A {\bf 66} 052103
\bibitem{49}
Chen F,  Klimchitskaya G L,
Mos\-te\-pa\-nen\-ko V M and Mohideen U 2006
{\it Phys. Rev. Lett.}  {\bf 97} 170402
\bibitem{50}
Rosenbaum T F, Milligan R F, Paalanen M A, Thomas G A,
Bhatt R N and Lin W 1983
{\it Phys. Rev.} B {\bf 27} 7509
\bibitem{51}
Dai P, Zhang Y and Sarachik M P 1991
{\it Phys. Rev. Lett.}  {\bf 66} 1914
\bibitem{52}
Winkler M T, Recht D, Sher M-J, Said A J, Mazur E
and Aziz M J 2011
{\it Phys. Rev. Lett.}  {\bf 106} 178701
\bibitem{53}
Ertekin E, Winkler M T, Recht D, Said A J, Aziz M J,
Buonassisi T and Grossman J C 2012
{\it Phys. Rev. Lett.}  {\bf 108} 026401
\bibitem{54}
Klein T, Achatz P, Kacmarcik J, Marcenat C, Gustafsson F,
Marcus J, Bustarret E, Pernot J, Omnes F, Sernelius Bo E,
Persson C, Ferreira da Silva A and Cytermann C 2007
{\it Phys. Rev.} B {\bf 75} 165313
\bibitem{55Si}
Palik E D (ed) 1991 {\it Handbook of Optical Constants of Solids},
vol 2 (New York: Academic)
\bibitem{56}
Lambrecht A, Pirozhenko I, Duraffourg L and
Andreucci Ph 2007
{\it Europhys. Lett.} {\bf 77} 44006
\bibitem{55Au}
Palik E D (ed) 1985 {\it Handbook of Optical Constants of Solids},
vol 1 (New York: Academic)
\bibitem {57}
Bimonte G 2011
{\it Phys. Rev.} A {\bf 83} 042109
\bibitem{58}
Bergstr\"{o}m L 1997
{\it Adv. Coll. Interface Sci.} {\bf 70} 125
\bibitem{59}
Ortolani M, Lupi S, Baldassarre L, Schade U, Calvani P, Takano Y,
Nagao M, Takenouchi T and Kawarada H 2006
{\it Phys. Rev. Lett.}  {\bf 97} 097002

\end{thebibliography}
\end{document}